\documentclass[preprint,aps,superscriptaddress,showpacs,floatfix]{revtex4}
\usepackage{graphicx}
\begin{document}
\title{Pentaquark $\Theta^+$ production from the reaction
$\gamma p\to\pi^+K^-\Theta^+$}
\bigskip
\author{W. Liu}
\affiliation{Cyclotron Institute and Physics Department, Texas A$\&$M
University, College Station, Texas 77843-3366, USA}
\author{C. M. Ko}
\affiliation{Cyclotron Institute and Physics Department, Texas A$\&$M
University, College Station, Texas 77843-3366, USA}
\author{V. Kubarovsky}
\affiliation{Rensselaer Polytechnic Institute, Troy, New York 12180-3590
and Thomas Jefferson National Accelerator Facility, Newport News, Virginia
23606}

\date{\today}

\begin{abstract}

The cross section for $\Theta^+$ production from the reaction
$\gamma p\to\pi^+K^-\Theta^+$, which was observed in the CLAS
experiment at the Jefferson National Laboratory, is evaluated in 
a hadronic model that includes couplings of $\Theta^+$ to both 
$KN$ and $K^*N$. With their coupling constants determined from 
the empirical $\pi NN(1710)$ and $\rho NN(1710)$ coupling constants
using the SU(3) symmetry, the cross section for this reaction has been 
evaluated by taking $\Theta^+$ to have spin 1/2 and isospin 0 but 
either positive or negative parity. We find that the cross section 
is 10-15 nb if $\Theta^+$ has positive parity as predicted by 
the chiral soliton model. The cross section is reduced by more than 
a factor of 10 if $\Theta^+$ has negative parity as given by 
lattice QCD studies. For both parities, the differential distribution 
peaks at small negative four momentum transfer as expected from the
dominating $t$-channel kaon-exchange diagram that involves only
the coupling of $\Theta^+$ to $KN$.

\end{abstract}


\maketitle

\section{introduction}

One of the most exciting recent experiments in hadron spectroscopy is the
detection of a narrow baryon state from the invariant mass spectrum of
$K^+n$ or $K^0p$ in nuclear reactions induced by photons
\cite{nakano,stepanyan,kubarovsky,barth} or kaons \cite{barmin}.
The extracted mass of about 1.54 GeV and width of less than 21-25 MeV
are consistent with those of the pentaquark baryon $\Theta^+$ consisting
of $uudd\bar s$ quarks and predicted in the chiral soliton model
\cite{diakonov}. Its existence has also been verified in the Skyrme 
model \cite{prasz,polyakov,walliser,jennings,borisyuk,itzhaki}, 
the constituent quark model \cite{riska,lipkin,jaffe}, the chiral quark model
\cite{hosaka,glozman}, the QCD sum rules \cite{zhu,matheus,sugiyama}, and 
the lattice QCD \cite{sasaki,ciskor}. Although most models predict
that $\Theta^+$ has spin 1/2 and isospin 0, their predictions on  
$\Theta^+$ parity vary widely.  While the soliton model gives a positive 
parity and the lattice QCD studies favors a negative parity, the quark 
model can give either positive or negative parities, depending on whether 
quarks are correlated or not.  Since the quantum numbers of the detected 
$\Theta^+$ are not yet determined in experiments, studies have
therefore been carried out to predict its decay branching ratios based
on different assignments of the $\Theta^+$ quantum numbers \cite{carl,ma}.

To evaluate the cross sections for $\Theta^+$ production from these reactions,
we have employed a hadronic model that is based on gauged SU(3) flavor
symmetric Lagrangians with the photon introduced as a U$_{\rm em}$(1)
gauged particle \cite{liuko}. The symmetry breaking effects are taken
into account phenomenologically by using empirical hadron masses and
coupling constants as well as empirical form factors. For the reaction 
$\gamma p\to\bar K^0\Theta^+$, in which $\Theta^+$ was detected in the 
SAPHIR experiment at Bonn University's ELSA accelerator \cite{barth}, 
the predicted cross section reaches a value of about 40 nb at photon 
energy $E_\gamma\sim 3.5$ GeV if the parity of $\Theta^+$ is taken to 
be positive. This value is about an order-of-magnitude smaller than the 
300 nb measured in the experiment. On the other hand, the reaction 
$\gamma n\to K^-\Theta^+$, which corresponds to the one seen in the 
LEPS experiment at SPring-8 \cite{nakano} and the CLAS experiment at 
the Thomas Jefferson National Laboratory (JLab) \cite{stepanyan}, 
is predicted to have a peak value of about 280 nb at $E_\gamma\sim 2.2$ 
GeV, comparable to that obtained in Ref.\cite{nam} using similar 
hadronic Lagrangians. The much larger cross section for the reaction
$\gamma n\to K^-\Theta^+$ than that for the reaction
$\gamma p\to\bar K^0\Theta^+$ is due to coupling of the photon to 
the virtual $K^-$ accompanied with the produced $\Theta^+$ 
\cite{liu}, which is absent in the latter reaction. It was also pointed 
out in Ref.\cite{liu} that if $\Theta^+$ is allowed to couple to $K^*N$ 
with a coupling constant similar to the $KN\Theta$ coupling and if one
further includes the photon anomalous parity interactions with kaons,
which are responsible for the decay of $K^*$ to $K\gamma$, then the
cross section for the reaction $\gamma p\to\bar K^0\Theta^+$ is
increased to about 350 nb, comparable to that quoted in the
SAPHIR experiment, although that for the reaction
$\gamma n\to K^-\Theta^+$ is only somewhat enhanced.
In these studies, photon interactions with the anomalous magnetic moment of
nucleons are, however, not considered. Including this contribution
and using empirical form factors extracted from photoproduction of
lambda from protons, the cross sections for these reactions have
been evaluated in Ref.\cite{oh} and were again found to be
a few hundred nb except that the one on protons is somewhat larger
than the one on neutrons. In both Ref.\cite{nam} and Ref.\cite{oh}, it 
was further found that the $\Theta^+$ production cross section is
significantly reduced if its parity is negative.

Besides the reactions $\gamma p\to\bar K^0\Theta^+$ and
$\gamma n\to K^-\Theta^+$, the $\Theta^+$ was also observed in the
reaction $\gamma p\to\pi^+K^-\Theta^+$ by the CLAS collaboration at JLab
\cite{kubarovsky}. It was suggested in Ref.\cite{kubarovsky} that this 
process is likely to result from the decay of $\bar K^{*0}$ in the reaction 
$\gamma p\to\bar K^{*0}\Theta^+$. Since this cross section has not
been considered in previous studies, we shall evaluate it in this paper
using the hadronic Lagrangians introduced in Refs.\cite{liuko,liu}
but with improved considerations of the $\Theta^+$ coupling constants
and the form factors at strong interaction vertices. Our results show 
that the magnitude of the cross section is sensitive to the parity of 
$\Theta^+$ and is thus useful in the experimental determination of 
$\Theta^+$ parity.

\section{$\Theta^+$ production from the reaction 
$\gamma p\to\pi^+K^-\Theta^+$}

\begin{figure}[ht]
\includegraphics[width=4in,height=2in,angle=0]{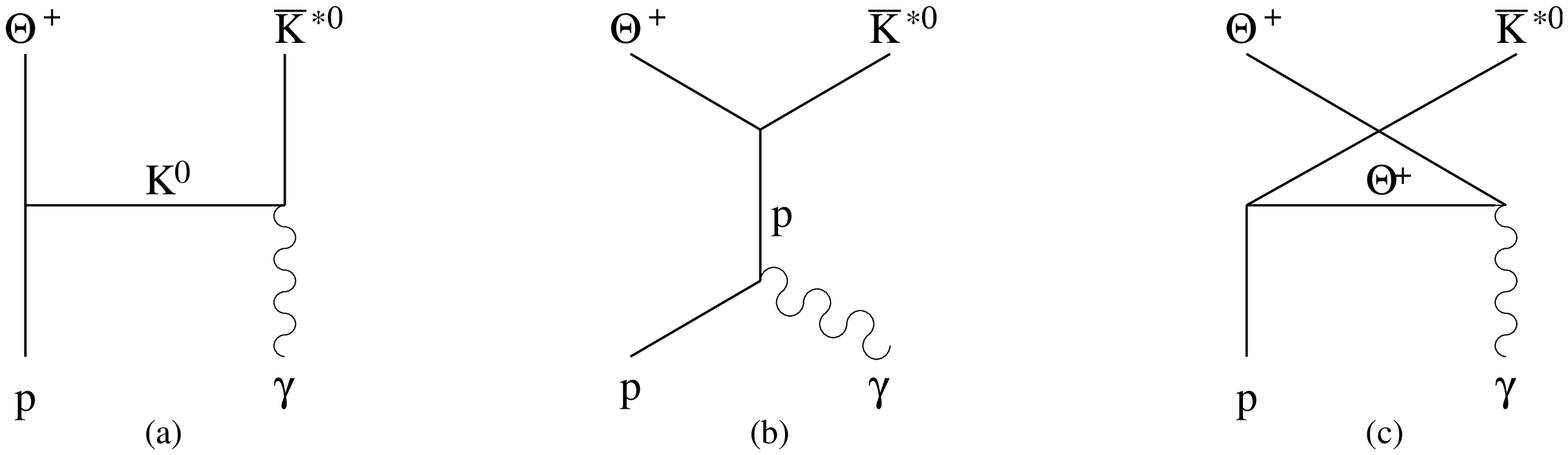}
\caption{Diagrams for $\Theta^+$ production from the reaction
$\gamma p\to\bar K^{*0}\Theta^+$.}
\label{diagram1}
\end{figure}

The diagrams that contribute to the reaction $\gamma p\to\bar K^{*0}\Theta^+$
are shown in Fig.\ref{diagram1}. If we take $\Theta^+$ to have spin 1/2, 
isospin 0, and positive parity, then the following interaction Lagrangians
are needed to evaluate the amplitudes for these diagrams:
\begin{eqnarray}\label{lagrangian}
{\cal L}_{KN\Theta}&=&ig_{KN\Theta}\bar N\gamma_5\Theta K+{\rm h.c.},
\nonumber\\
{\cal L}_{K^*N\Theta}&=&g_{K^*N\Theta}\bar N\gamma_\mu\Theta
K^{*\mu}+{\rm h.c.},\nonumber\\
{\cal L}_{\gamma NN}&=&-e\bar N\left\{\gamma_\mu\frac{1+\tau_3}{2}
-\frac{1}{4m_N}\left[\kappa_p+\kappa_n
+\tau_3(\kappa_p-\kappa_n)\right]\sigma_{\mu\nu}\partial^\nu\right\} 
A^\mu N,\nonumber\\
{\cal L}_{\gamma\Theta\Theta}&=&-e\bar\Theta\gamma_\mu\Theta A^\mu,\nonumber\\
{\cal L}_{\gamma KK^{*}}&=&g_{\gamma KK^*}\epsilon_{\alpha\beta\mu\nu}
\partial^\alpha A^\beta[\partial^\mu\bar K^{*\nu}K
+\bar K\partial^\mu K^{*\nu}].
\end{eqnarray}
In the above, $A$ denotes the photon field while $N$, $K$, and $K^*$
are the isospin doublet nucleon, kaon, and vector kaon fields, 
respectively; $\gamma_\mu$ and $\gamma_5$ are Dirac matrices while
$\tau_3$ is the Pauli spin matrix; and $\epsilon_{\alpha\beta\mu\nu}$ 
represents the antisymmetric tensor with the usual convention 
$\epsilon_{0123}=1$. 

The coupling constant $g_{KN\Theta}$ can in principle be determined
from the width $\Gamma_\Theta$ of $\Theta^+$. Unfortunately, $\Gamma_\Theta$
is not well known due to limitation of experimental resolutions. In our 
previous studies \cite{liuko,liu}, we have taken the experimental
upper limit of $\Gamma_\Theta=20$ MeV and obtained $g_{KN\Theta}=4.4$ using 
\begin{eqnarray}\label{width}
\Gamma_\Theta=\frac{g^2_{KN\Theta}}{2\pi}
\frac{k(\sqrt{m^2_N+k^2}-m_N)}{m_\Theta},
\end{eqnarray}
where $k$ is the momentum of nucleon with mass $m_N$ or kaon with mass
$m_K$ in the rest frame of $\Theta^+$. In the present study, we use 
SU(3) symmetry to relate $g_{KN\Theta}$ to $g_{\pi NN_{10}}$ between 
the coupling of pentaquark $N_{10}$ ($uudd\bar u$, $uudd\bar d$) and 
$N\pi$, i.e., $g_{KN\Theta}=\sqrt{6}g_{\pi NN_{10}}$ \cite{kim}, 
with $g_{\pi NN_{10}}$ determined from the decay width of $N_{10}$ 
to $N\pi$, given by an expression similar to Eq.(\ref{width}) with 
$g_{KN\Theta}$, $m_\Theta$, and $m_K$ replaced by $\sqrt{3/2}$ times 
the $\pi NN_{10}$ coupling $g_{\pi NN_{10}}$, the $N_{10}$ mass 
$m_{10}$, and the pion mass $m_\pi$, respectively.

If we assume that the pentaquark $N_{10}$ is the $N(1710)$ in the 
particle data book \cite{particle}, we then obtain 
$g_{\pi NN_{10}}\sim 1.0$ and thus $g_{KN\Theta}\sim 2.5$ from
$\Gamma_{N(1710)\to N\pi}\sim 100\times 0.15=15$ MeV, where $100$ MeV 
is the total decay width of $N(1710)$ and 0.15 is the branching 
ratio for decaying to $N\pi$ \cite{particle}. However, it was pointed
out in Ref.\cite{kim} that $N(1710)$ cannot be a pure antidecuplet 
pentaquark baryon as it then cannot have the large decay branching ratio 
to $\Delta\pi$ seen empirically. Following Ref.\cite{kim}, we 
assume that antidecuplet pentaquark baryons are mixed ideally with
octet pentaquark baryons. In this case, the $\Theta^+$ remains 
a pure antidecuplet pentaquark baryon but the mixed pentaquark baryons
($\sqrt{2}N_{10}-N_8)/\sqrt{3}$, where $N_8$ 
($udds\bar s$, $uuds\bar s$) are octet pentaquark baryons, can be 
identified with the $N(1710)$ in the particle data book.
This then leads to the coupling constant 
$g_{\pi NN_{10}}\sim 1.0\times\sqrt{3/2}\sim 1.2$ and thus 
$g_{KN\Theta}=3g_{\pi NN_{10}}\sim 3.0$. According to Eq.(\ref{width}), 
this would give a $\Theta^+$ width of $\Gamma_\Theta\sim 9$ MeV, which
is smaller than that seen in photonucleon reactions 
\cite{nakano,stepanyan,kubarovsky,barth} but is larger than that 
in kaon-nucleon reactions \cite{barmin}.

The coupling constant $g_{K^*N\Theta}$ between $\Theta^+$ and $NK^*$ 
can be similarly determined from the SU(3) relation 
$g_{K^*N\Theta}=3g_{\rho NN(1710)}$. Since the rho meson has a large
decay width, the decay width of $N_{10}$ to $N\rho$ is given 
differently from that for its decay to $N\pi$, i.e.,  
\begin{eqnarray}
\Gamma_{N(1710)\to N\rho}&=&\frac{3g^2_{\rho NN(1710)}}{4\pi}\int dm~M(m)
\frac{k}{m_{N(1710)}}\nonumber\\
&&\times\left[\sqrt{m_N^2+k^2}-3m_N
+\frac{\sqrt{m^2+k^2}}{m^2}(m^2_{N(1710)}-m^2-m^2_N)\right],
\end{eqnarray}
where $k$ is the three momentum of nucleon or rho meson 
with mass $m$ in the rest frame of $N(1710)$. In the above,  
\begin{equation}
M(m)=\frac{\alpha\Gamma(m)}{2\pi[(m-m_\rho)^2+\Gamma^2(m)/4]}
\end{equation}
is the rho meson mass distribution with mass-dependent width
$\Gamma(m)=\Gamma_\rho(k/k_\rho)^3(m_\rho/m)^3$,
where $\Gamma_\rho=150$ MeV is the empirical width of rho meson when
its mass is $m_\rho=770$ MeV, and $k_\rho=\sqrt{m_\rho^2/4-m_\pi^2}$ and
$k=\sqrt{m^2/4-m_\pi^2}$.  
The constant $\alpha=1.256$ is introduced to normalize the rho mass 
distribution, i.e., $\int dm~M(m)=1$. Using the empirical value
$\Gamma_{N(1710)\to N\rho}\sim 15$ MeV, we obtain $g_{K^*N\Theta}\sim 1.8$.
As the sign of $g_{K^*N\Theta}$ relative to
that of $g_{KN\Theta}$ cannot be fixed by SU(3) symmetry, we shall
consider both signs for the coupling constant $g_{K^*N\Theta}=\pm 1.8$ 
as well as for $g_{K^*N\Theta}=0$. 

For photon coupling to nucleon, we include also its interaction
with the anomalous magnetic moment of nucleons with empirical values
of $\kappa_p=1.79$ and $\kappa_n=-1.91$. Since the anomalous magnetic 
moment of $\Theta^+$ is not known, we neglect its coupling to photon. 

The coupling constant $g_{\gamma KK^*}$ denotes the photon anomalous 
parity interaction with kaons and has the dimension of inverse of 
energy. Its value is $g_{\gamma K^0K^{*0}}=0.388$ GeV$^{-1}$ using the 
decay width $\Gamma_{K^{*0}\to K^0\gamma}=0.117$ MeV of $K^{*0}$ 
to kaon and photon \cite{liu}. Although the sign of $g_{\gamma KK^*}$ 
relative to other coupling constants in the interaction Lagrangians is 
not known either, it is not relevant for our study as both constructive 
and destructive interferences among the three diagrams in 
Fig. \ref{diagram1} are automatically taken into account by using 
different signs for the coupling constant $g_{K^*N\Theta}$.

The amplitudes for the three diagrams shown in Fig. \ref{diagram1}  
for the reaction $\gamma p\to\bar K^{*0}\Theta^+$ are then given, 
respectively, by
\begin{eqnarray}\label{amplitude}
{\cal M}_t&=&ig_{\gamma K^0K^{*0}}g_{KN\Theta}
\bar\Theta(p_3)\gamma_5\frac{\epsilon_{\alpha\beta\mu\nu}
p_2^\alpha\epsilon_2^\beta p_4^\mu\epsilon_4^\nu}{t-m^2_K}p(p_1),\nonumber\\
{\cal M}_s&=&-eg_{K^*N\Theta}\bar\Theta(p_3)\epsilon_4\cdot\gamma
\frac{(p_1+p_2)\cdot\gamma+m_N}{s-m^2_N}
\left[1+\frac{\kappa_p}{2m_N}p_2\cdot\gamma\right]
\epsilon_2\cdot\gamma~p_(p_1)\nonumber\\
{\cal M}_u&=&-eg_{K^*N\Theta}\bar\Theta(p_3)\epsilon_2\cdot\gamma
\frac{(p_1-p_4)\cdot\gamma+m_\Theta}{u-m^2_\Theta}
\epsilon_4\cdot\gamma~p(p_1).
\end{eqnarray}
In the above, we have introduced the usual Mandelstan variables
$s=(p_1-p_2)^2$, $t=(p_1-p_3)^2$, and $u=(t_1-t_4)^2$, with $p_1$, 
$p_2$, $p_3$, and $p_4$ denoting the momenta of proton, photon,
$\Theta^+$, and $K^{*0}$, respectively. The polarization vectors of 
photon and $K^{*0}$ are given, respectively, by $\epsilon_2$ 
and $\epsilon_4$.   

To take into account the internal structure of hadrons, form factors 
are needed at strong interaction vertices.  In our previous studies
of $\Theta^+$ production in the reactions $\gamma N\to\bar K\Theta$, 
we have used the same form factor 
$F({\bf q}^2)=\Lambda^2/(\Lambda^2+{\bf q}^2)$, where ${\bf q}$ is the
three momentum of photon in the center of mass frame, for all
amplitudes in order to keep the total amplitude gauge invariant.
The cutoff parameter $\Lambda=0.75$ GeV used in these
studies is obtained from fitting the empirical charmed hadron 
production cross section in photon-proton reactions at 6 GeV with
similar interaction Lagrangians based on SU(4) flavor symmetry with 
empirical hadron masses and coupling constants \cite{liu2}. Although this 
form factor suppresses the growth of total cross section with increasing 
center of mass energy, it does not damp sufficiently the increasing
contribution of $t-$channel diagram to the differential cross section 
at large four momentum transfer. This can be improved by using 
a form factor for the $t-$channel amplitude that depends on the 
momentum of the exchanged particle instead of the momentum of the photon.
Since the $t-$channel amplitude is gauge invariant by itself, the
total amplitude remains gauge invariant even the form factor for the 
$t-$channel amplitude is different from that for the $s-$ and
$u-$channel amplitudes.In the present study, we follow, however, 
the method of Ref.\cite{janssen} for photoproduction of lambda from 
protons by introducing different covariant form factors
for the $s-$, $t-$, and $u-$channel amplitudes, and they are
given by   
\begin{eqnarray}
F(x)=\frac{\Lambda^4}{\Lambda^4+(x-m_x^2)^2},
\label{form}
\end{eqnarray}
where $x=s$, $t$, and $u$ with corresponding masses $m_x=m_N$, $m_K$, 
and $m_\Theta$ of the off-shell particles at strong interaction 
vertices. The cutoff parameter $\Lambda$ 
characterizes the off-shell momentum above which hadron internal 
structure becomes important, and its value will be determined 
empirically.  Although the $t-$channel amplitude $M_t$ after including 
the form factor $F(t)$ remains gauge invariant, adding form factors 
$F(s)$ and $F(t)$ respectively to the $s-$ and $t-$channel amplitudes 
$M_s$ and $M_u$ leads to terms that violate the gauge invariance. 
The gauge violation can, however, be removed by introducing a contact 
term in the interaction Lagrangian that gives an additional amplitude 
of the form
\begin{eqnarray}\label{contact}
{\cal M}_c&=&-2eg_{K^*N\Theta}\bar\Theta(p_3)\epsilon_4\cdot\gamma
\left[\frac{\epsilon_2\cdot p_1}{s-m^2_N}(\hat F-F(s))
+\frac{\epsilon_2\cdot p_3}{u-m^2_\Theta}(\hat F-F(u))\right]p(p_1),
\end{eqnarray}
with $\hat F=1$ \cite{ohta}, $a_sF(s)+a_uF(u)$ ($a_s+a_u=1$)
\cite{haberzettl}, or ${\hat F}=F(s)+F(u)-F(s)F(u)$ \cite{workman}.
Here, we adopt the last $\hat F$ in order to maintain both
crossing symmetry and the pole structure of original amplitude.
To determine the value of cutoff parameter $\Lambda$, we again use similar
interaction Lagrangians based on the SU(4) flavor symmetry with empirical
masses and coupling constants to study charmed hadron production 
from photon-proton reactions at center-of-mass energy of 6 GeV \cite{liu2}.
Comparisons with available experimental data gives
$\Lambda=0.8$ GeV, which is similar to the soft form factor considered 
in Ref. \cite{janssen} for describing experimental data on lambda
production in photoproton reactions.

The resulting differential cross section for the reaction 
$\gamma p\to\bar K^{*0}\Theta^+$ is then
\begin{eqnarray}\label{dsdt}
\frac{d\sigma_{\gamma p\to K^{*0}\Theta^+}}{dt}&=&
\frac{1}{256\pi sp^2_i}|F(t)M_t+F(s)M_s+F(u)M_u+M_c|^2,
\end{eqnarray}
where $p_i$ is the magnitude of the three momenta of initial-state 
particles in the center-of-mass frame, i.e., $p_i=(s-m_N^2)/2\sqrt{s}$.

Since $\bar K^{*0}$ can decay into either $\pi^+K^-$ or $\pi^0\bar K^0$
with a branching ratio of 2 to 1, the differential cross section for
the reaction $\gamma p\to\pi^+K^-\Theta^+$ is thus 2/3 of that given 
by Eq.(\ref{dsdt}). 

\section{results}

\subsection{positive parity $\Theta^+$}

\begin{figure}[ht]
\includegraphics[width=3.25in,height=3.25in,angle=-90]{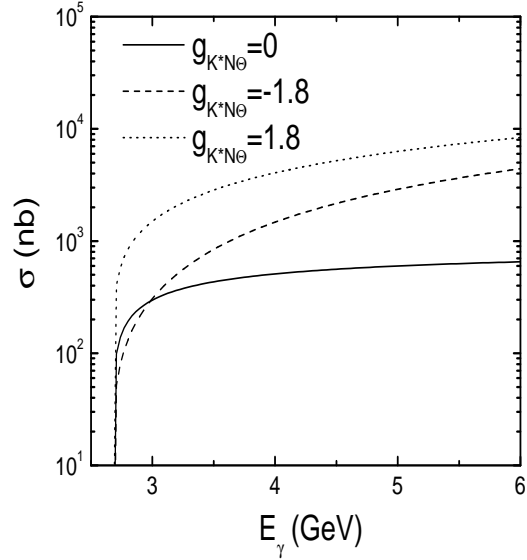}
\caption{Total cross section for the production of positive parity 
$\Theta^+$ from the reaction $\gamma p\to\pi^+K^-\Theta^+$ as a 
function of photon energy obtained without form factors and for 
the coupling constant $g_{K^*N\Theta}=1.8$ (dotted curves), 0 
(solid curves), and -1.8 (dashed curves).}
\label{cross1}
\end{figure}

We first show in Fig. \ref{cross1} the total cross section for the 
production of a positive parity $\Theta^+$ in the reaction 
$\gamma p\to\pi^+K^-\Theta^+$ obtained without form factors 
but with different values of 1.8 (dotted curves), 0 (solid curves), 
and -1.8 (dashed curves) for the coupling constant $g_{K^*N\Theta}$.
It is seen that the cross section with $g_{K^*N\Theta}=0$,
corresponding to neglect of $s-$ and $u-$channels diagrams (b) and (c) 
of Fig. \ref{diagram1}, generally has the smallest value. This implies that
$s-$ and $u-$channel contributions due to $K^*N\Theta$ coupling are
most important when we neglect the internal structure of hadrons.

\begin{figure}[ht]
\includegraphics[width=3.25in,height=3.25in,angle=-90]{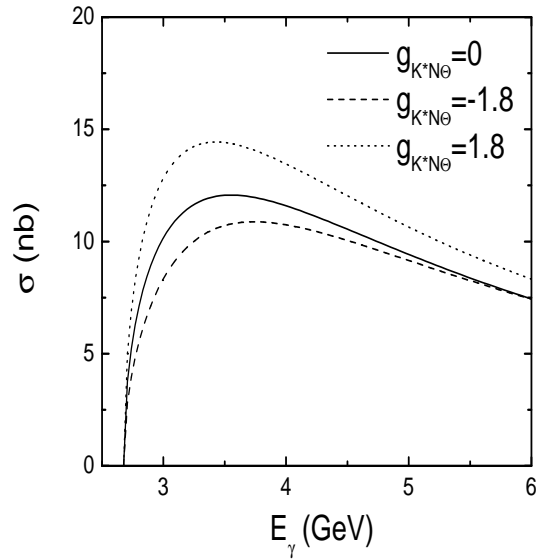}
\caption{Same as Fig.\ref{cross1} with form factors.}
\label{cross2}
\end{figure}

Including form factors at strong interaction vertices according to the
method described in the previous section, the resulting total cross sections 
are shown in Fig. \ref{cross2} for the coupling constant 
$g_{K^*N\Theta}=1.8$ (dotted curves), 0 (solid curves), and -1.8 
(dashed curves). In this case, the cross sections are
more than two orders of magnitude smaller than corresponding ones 
shown in Fig. \ref{cross1} without form factors. Furthermore, unlike
the case without form factors, the dependence of the cross section on 
the value of $g_{K^*N\Theta}$ is small. This result thus shows that the 
effect of form factors on $s$ and $u$ channels are much stronger than on 
the $t$ channel, which gives the dominant contribution when the hadron
internal structure is taken into account. Depending on the value of
$g_{K^*N\Theta}$, the total cross section obtained with form factors 
has a peak value of 10-15 nb at photon energy $E_\gamma\sim 3.5$ GeV.

\begin{figure}[ht]
\includegraphics[width=3.5in,height=3.5in,angle=-90]{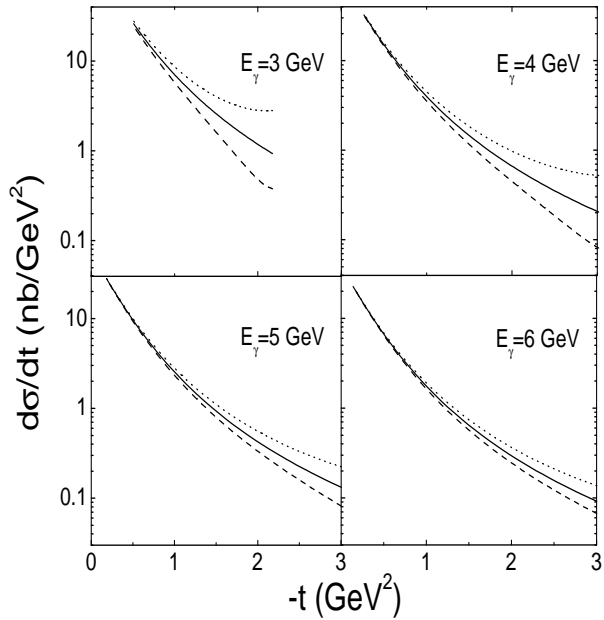}
\caption{Differential cross section for production of a positive 
parity $\Theta^+$ from the reaction $\gamma p\to\pi^+K^-\Theta^+$ 
at $E_\gamma$=3, 4 , 5, and 6 GeV obtained with form factors and 
for the coupling constant $g_{K^*N\Theta}=1.8$ (dotted curves), 
0 (solid curves), and -1.8 (dashed curves).}
\label{across1}
\end{figure}

In Fig. \ref{across1}, we show the differential cross section for
producing a positive parity $\Theta^+$ in the reaction 
$\gamma p\to\pi^+K^-\Theta^+$ as a function of negative four momentum
transfer $-t$ for photon energies of $E_\gamma=$3, 4, 5, and 6 GeV and 
using different values of 1.8 (dotted curves), 0 (solid curves), and 
-1.8 (dashed curves) for the coupling constant $g_{K^*N\Theta}$.
In all cases, the differential cross section peaks at small negative 
four momentum transfer as expected from the dominant $t-$channel 
diagram in Fig. \ref{diagram1}.

\subsection{negative parity $\Theta^+$}

We have assumed in the above that the parity of $\Theta^+$ is
positive as predicted by the chiral soliton model \cite{diakonov}. 
On the other hand, results from lattice QCD calculations have indicated 
that the mass of $\Theta^+$ observed in experiments is consistent 
with that of a negative parity pentaquark baryon with spin 1/2 and 
isospin 0 \cite{ciskor,sasaki}. In this case, the interaction
Lagrangians involving $\Theta^+$ are given by 
\begin{eqnarray}
{\cal L}_{KN\Theta}&=&g_{KN\Theta}\bar N\Theta K+{\rm h.c.},
\nonumber\\
{\cal L}_{K^*N\Theta}&=&ig_{K^*N\Theta}\bar
N\gamma_5\gamma_\mu\Theta K^{*\mu}+{\rm h.c.}.
\end{eqnarray}
This changes the expression for the $\Theta^+$ width shown in 
Eq.(\ref{width}), i.e., the $-m_N$ is replaced by $m_N$. As a result, 
the coupling constant $g_{KN\Theta}$ is reduced by about a factor of 7, 
i.e., $g_{KN\Theta}=0.42$ if the $\Theta^+$ width is taken to be 9 MeV. 
Since $N(1710)$ is known to have positive parity, it 
could not be in the same multiplet as $\Theta^+$ if the latter
has negative parity.  The SU(3) relation used previously for relating 
$g_{K^*N\Theta}$ to $g_{\rho NN(1710)}$ is then not applicable. Instead, 
we assume that the ratio $g_{K^*N\Theta}/g_{KN\Theta}$ remains the 
same whether $\Theta^+$ has positive or negative parity. This leads to 
$g_{K^*N\Theta}=0.25$ for negative parity $\Theta^+$. 

The amplitudes for the three diagrams in Fig. \ref{diagram1} for producing
a negative parity $\Theta^+$ are similar to those in Eq.(\ref{amplitude})
for positive parity $\Theta^+$ except that the factor $i\gamma_5$ 
should be inserted after $\bar\Theta(p_1)$ in all three amplitudes.
The $i\gamma_5$ is also needed in the amplitude shown in Eq.(\ref{contact})
due to the contact interaction introduced for canceling the gauge
violating terms after form factors are included at strong
interaction vertices. The cross section formula for the reaction 
$\gamma p\to\pi^+K^-\Theta^+$ with negative parity $\Theta^+$ remains
to be 2/3 of Eq.(\ref{dsdt}).

\begin{figure}[ht]
\includegraphics[width=3.25in,height=3.25in,angle=-90]{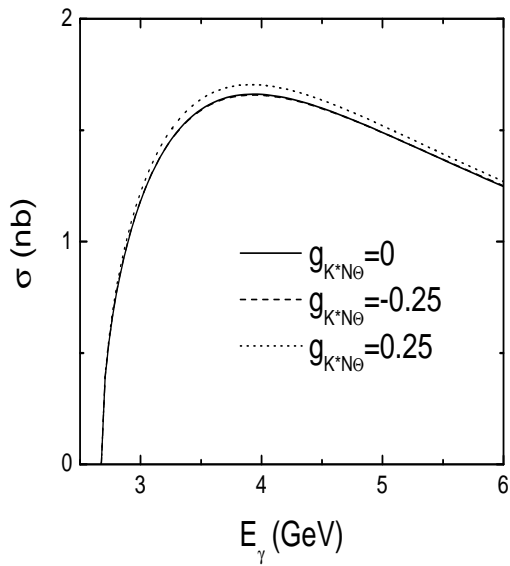}
\caption{Total cross section for the production of negative parity 
$\Theta^+$ from the reaction $\gamma p\to\pi^+K^-\Theta^+$ as a 
function of photon energy for the coupling constant 
$g_{K^*N\Theta}=0.25$ (dotted curves), 0 (solid curves), and -0.25 
(dashed curves).}
\label{cross3}
\end{figure}

The total cross section for producing a negative parity $\Theta^+$
from the reaction $\gamma p\to\pi^+K^-\Theta^+$ is shown in 
Fig.\ref{cross3} for the three values of 0.25 (dotted curves), 0 
(solid curves), and -0.25 (dashed curves) for $g_{K^*N\Theta}$. 
Form factors at strong interaction vertices are taken to have the same
form shown in Eq.(\ref{form}) with cutoff parameter $\Lambda=0.8$ 
GeV as in the case of positive parity $\Theta^+$. It is seen that 
these cross sections are insensitive to the value of $g_{K^*N\Theta}$ 
as the dominant contribution is from the $t-$channel diagram involving 
only $g_{KN\Theta}$. Their magnitude is more than an order-of-magnitude 
smaller than that for producing a positive parity $\Theta^+$. 
The peak value is now only about 1.5 nb.

\begin{figure}[ht]
\includegraphics[width=3.25in,height=3.25in,angle=-90]{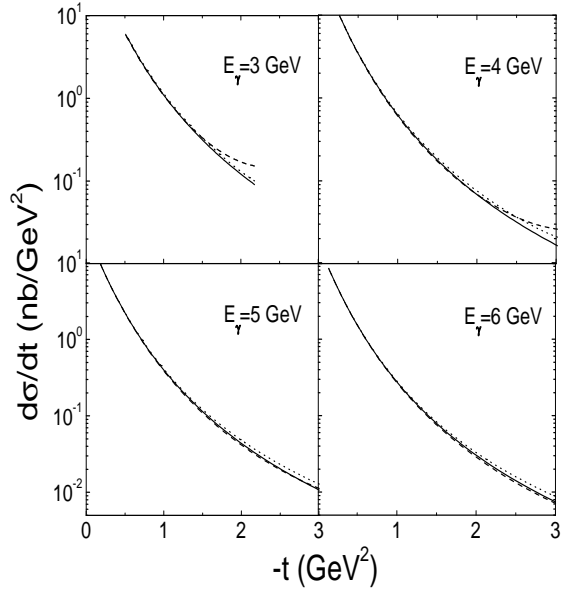}
\caption{Differential cross section for production of a negative
parity $\Theta^+$ from the reaction $\gamma p\to\pi^+K^-\Theta^+$ 
at $E_\gamma$=3, 4 , 5, and 6 GeV obtained with form factors and for 
the coupling constant $g_{K^*N\Theta}=0.25$ (dotted curves), 
0 (solid curves), and -0.25 (dashed curves).}
\label{across2}
\end{figure}

In Fig. \ref{across2}, the differential cross section for producing a
negative parity $\Theta^+$ as a function of negative four momentum
transfer $-t$ is shown for four photon energies of $E_\gamma=3$, 4, 5, 
and 6 GeV with the coupling constant $g_{K^*N\Theta}$ having values 
of -0.25 (dashed curves), 0 (solid curves), and 0.25 (dotted curves). 
As in the production of positive parity $\Theta^+$, the differential
cross section for producing a negative parity $\Theta^+$ in the
reaction $\gamma p\to\pi^+K^-\Theta^+$ always peaks 
at small negative four momentum transfer.

\section{discussions}

\begin{figure}[ht]
\includegraphics[width=1.5in,height=2in,angle=0]{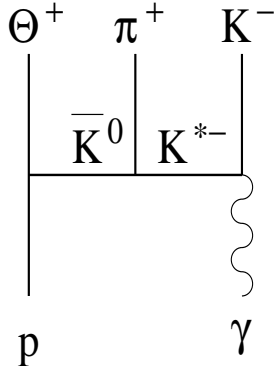}
\caption{Diagram for $\Theta^+$ production from the reaction
$\gamma p\to\pi^+K^-\Theta^+$ via $K^{*-}$ exchange.}
\label{diagram2}
\end{figure}

The reaction $\gamma p\to\pi^+K^-\Theta^+$ can also proceed via the
exchange of $K^{*-}$ as shown in Fig. \ref{diagram2} besides 
through the decay of $\bar K^{*0}$ shown in Fig. \ref{diagram1}. 
This process involves an additional strong interaction vertex $K^*K\pi$ 
compared to those in Fig.\ref{diagram1}. Although the coupling constant
$g_{\pi KK^*}=3.28$, determined from the decay width 
$\Gamma_{K^*\to K\pi}=50.8$ MeV of $K^*$, is not small, the additional
form factor would reduce its contribution significantly compared to
that from the diagrams shown in Fig. \ref{diagram1}. Similar
considerations for the reactions $\gamma n\to\pi^0K^-\Theta^+$ 
and $\gamma n\to\pi^-\bar K^0\Theta^+$ have shown that the
contribution from the $K^*$ exchange is negligible
(about a few \%) compared to that from the $K^*$ decay \cite{liu}. 
It is thus safe to neglect its contribution.

The coupling constant $g_{KN\Theta}$ used in present study is determined 
using SU(3) symmetry from the coupling constant $g_{\pi NN(1710)}$, 
which can be extracted from the empirical decay width of $N(1710)$ 
to $N\pi$. The resulting value leads to a $\Theta^+$ decay
width of about 9 MeV, which is between those seen in photonucleon
experiments \cite{nakano,stepanyan,kubarovsky,barth} and 
in kaon-nucleon reactions \cite{barmin}. On the other hand, 
if $\Theta^+$ width is less than 1 MeV as suggested in 
Refs.\cite{arndt,haid} based on analyses of $K^+n$ scattering data, 
the cross sections shown in the above, which are dominated by 
$t-$channel contribution due to the $KN\Theta$ coupling,
would be reduced significantly as it is then proportional to the 
square of the coupling constant $g_{KN\Theta}$, similar to 
the width of $\Theta^+$ as shown in Eq.(\ref{width}).  

The calculated cross section is also sensitive to the value of cutoff 
parameter used in the form factors as shown by Figs.\ref{cross1} and
\ref{cross2} for results without and with form factors. If we use a harder 
form factor with cutoff parameter $\Lambda=1.6$ GeV, which is twice the
value used in present study, the resulting cross sections would 
increase by more than an order of magnitude. To have a 
reliable prediction for the cross section for the reaction 
$\gamma p\to\pi^+K^-\Theta^+$ as well as those for other 
$\Theta^+$ production reactions thus requires a good knowledge 
on both the coupling constants $g_{KN\Theta}$ as well as the form factors.

\section{summary}

Using interaction Lagrangians that involve couplings of $\Theta^+$ to
both $KN$ and $K^*N$, we have evaluated the cross section for its
production in the reaction $\gamma p\to\pi^+K^-\Theta^+$.  
The couplings constant $g_{KN\Theta}$ and $g_{K^*N\Theta}$
are determined via SU(3) relations from the empirical coupling constants 
$g_{\pi NN(1710)}$ and $g_{\rho NN(1710)}$ assuming that $N(1710)$ 
is an ideal mixture of antidecuplet and octet pentaquark baryon. 
With empirical cutoff parameter in the form factors, 
the cross section has been calculated for either 
positive or negative parity $\Theta^+$. We find that value of the 
cross section for producing a positive $\Theta^+$ is 10-15 nb
with dominating contribution from the $t-$channel kaon-exchange diagram
that involves only the $KN\Theta$ coupling. The cross section 
is reduced by an order of magnitude if $\Theta^+$ has negative parity.
In both cases, the differential cross section peaks at small negative 
four momentum transfer as expected from $t-$channel meson-exchange
processes.

\begin{acknowledgments}
V.K. is grateful to Maxim Polyakov for very helpful discussions.
This paper was based on work supported in part by the US National
Science Foundation under Grant No. PHY-0098805 and the Welch
Foundation under Grant No. A-1358.
\end{acknowledgments}

\end{document}